\documentclass[pra, twocolumn, superscriptaddress]{revtex4-2}

\usepackage{amsmath}
\usepackage{amssymb}
\usepackage{amsfonts}
\usepackage{amsthm}
\usepackage{mathtools}
\usepackage{subdepth}
\usepackage{esdiff}
\usepackage[normalem]{ulem}

\mathtoolsset{} 
\usepackage[T1]{fontenc} 
\usepackage[english]{babel}
\usepackage[utf8]{inputenc}
\usepackage{newtxtext, newtxmath} 
\usepackage{bm} 

\usepackage{dsfont}
\usepackage{graphicx}
\usepackage{xcolor}
\usepackage[colorlinks=true, allcolors=blue]{hyperref}
\usepackage{url}
\usepackage{orcidlink}

\newtheoremstyle{normal}
    {\parskip}
    {\parskip}
    {}
    {\parindent}
    {\itshape}
    {.}
    {.5em}
    {}
\makeatletter
    \renewcommand{\@upn}{} 
    \renewenvironment{proof}[1][\proofname]{
        \pushQED{\qed}%
        \normalfont \topsep0\p@\@plus0\p@\relax
        \trivlist
        \item\relax
        {\itshape\hspace{\parindent}%
        #1\@addpunct{.}}\hspace\labelsep\ignorespaces
    }{%
        \popQED\endtrivlist\@endpefalse
    }
\makeatother
\theoremstyle{normal}

\newtheorem{observation}{Observation}

\newcommand{\ie}{i.e., }

\def\leqs{\leqslant}

\def\hatPsi{\hat\Psi}

\DeclareMathOperator{\Cov}{Cov}
\newcommand{\matrixid}{ \mathds{1} }

\newcommand{\ket}[1]{ | #1 \rangle }

\newcommand{\ketbra}[2]{ | #1 \rangle \! \langle #2 | }
\newcommand{\projector}[1]{ \ketbra{#1}{#1} }
\newcommand{\braoket}[3]{ \langle #1 | #2 | #3 \rangle }

\newcommand{\expect}[1]{ \langle #1 \rangle }

\newcommand{\variance}[1]{ ( \Delta{#1} )^2 }

\newcommand{\ivariance}[1]{ ( \Delta{#1} )^{-2} }

\newcommand{\qfi}{ {{F}} }
\newcommand{\qfii}[1]{ {F}_{#1} }
\newcommand{\qfim}{ {\bm{F}} }

\newcommand{\rma}{{\rm{a}}}
\newcommand{\rmb}{{\rm{b}}}
\newcommand{\hatx}{{\hat{x}}}

\makeatletter
\@addtoreset{section}{mysection}
\makeatother

\makeatletter
\@addtoreset{equation}{myequation}
\makeatother

\makeatletter
\@addtoreset{figure}{myfigure}
\makeatother

\begin{document} 

\preprint{Preprint/1} 

\title{Differential magnetometry with partially flipped Dicke states}

\author{Iagoba Apellaniz\,\orcidlink{0000-0002-3688-3242}}
\affiliation{Department of Theoretical Physics, University of the Basque Country UPV/EHU,
P.O. Box 644, 48080 Bilbao, Spain}
\affiliation{EHU Quantum Center, University of the Basque Country UPV/EHU,
P.O. Box 644, 48080 Bilbao, Spain}
\author{Manuel Gessner\,\orcidlink{0000-0003-4203-0366}}
\affiliation{Departament de Física Teòrica, IFIC, Universitat de València, CSIC, Carrer del Dr.~Moliner 50, 46100 Burjassot (València), Spain} 
\author{Géza Tóth\,\orcidlink{0000-0002-9602-751X}}
\affiliation{Department of Theoretical Physics, University of the Basque Country  UPV/EHU,
P.O. Box 644, 48080 Bilbao, Spain}
\affiliation{EHU Quantum Center, University of the Basque Country UPV/EHU,
P.O. Box 644, 48080 Bilbao, Spain}
\affiliation{Donostia International Physics Center DIPC, Paseo Manuel de Lardizabal 4, 20018 San Sebasti\'an, Spain}
\affiliation{IKERBASQUE, Basque Foundation for Science, 48009 Bilbao, Spain}
\affiliation{HUN-REN Wigner Research Centre for Physics, P.O. Box 49, 1525 Budapest, Hungary}

\date{\today}

\begin{abstract}
    We study magnetometry of gradients and homogeneous background fields along the three orthogonal directions using two spatially separated spin ensembles.
    We derive trade-off relations for the achievable estimation precision of these parameters. 
    Dicke states, optimal for homogeneous field estimation, can be locally rotated into states sensitive to magnetic gradients by rotating the spins in one subensemble. 
    We determine bounds for the precision for gradient metrology in the three orthogonal directions as a function of the sensitivities to the homogenous field in those directions.
    The resulting partially flipped Dicke state saturates the bounds above, showing similar sensitivity in two directions but significantly reduced sensitivity in the third. 
    Exploiting entanglement between the two ensembles, this state achieves roughly twice the precision attainable by the best bipartite separable state, which is a product of local Dicke states.
    For small ensembles, we explicitly identify measurement operators saturating the quantum Cramér-Rao bound, while for larger ensembles, we propose simpler but suboptimal schemes. In both cases, the gradient is estimated from second moments and correlations of angular momentum operators. 
    Our results demonstrate how the metrological properties of Dicke states can be exploited for quantum-enhanced multiparameter estimation.
\end{abstract}

\maketitle

\section{Introduction}

In recent years, quantum metrology has been applied in a wide range of scenarios, from
atomic clocks \cite{Louchet-Chauvet2010Entanglement-assisted, Borregaard2013Near, Kessler2014Heisenberg} and
precision magnetometry \cite{Wasilewski2010Quantum,Eckert2006Differential, Wildermuth2006Sensing, Wolfgramm2010Squeezed, Koschorreck2011High,Vengalattore2007High, Zhou2010Precisely} to gravitational wave
detectors \cite{Schnabel2010Quantum,TheLIGOScientificCollaboration2011,Demkowicz-Dobrzanski2013Fundamental}. 
Concerning metrological applications, it has been found that spin-squeezed states of cold atoms and cold trapped ions make it possible to overcome the shot-noise limit in parameter estimation \cite{Kitagawa1993Squeezed,Wineland1994Squeezed,Sorensen2001Many-particle,Sorensen2001Entanglement,Hald1999Spin,Appel2009Mesoscopic,Wasilewski2010Quantum,Gross2010Nonlinear,Muessel2014Scalable}, while other quantum  states, such as Dicke states \cite{Lucke2011Twin,Krischek2011Useful,Lucke2014Detecting,Zou2018Beating,Chapman2023Long-Lived} and Greenberger-Horne-Zeilinger (GHZ) states \cite{Greenberger1990Bells,Sackett2000Experimental,Leibfried2004Toward,Monz201114-Qubit,Compact2021Pogorelov,Moses2023ARace-Track,Song201710-Qubit,Zhong201812Photon,Bao2024Schrodinger,Thomas2022Efficient} enable, at least theoretically, to reach the maximal Heisenberg scaling.

After extensive studies of the problem of estimating a single parameter, recently, increased attention is dedicated to multiparameter estimation problems \cite{Kolodynski2010Phase,Crowley2014Tradeoff,Monras2011Measurement, Vaneph2013Quantum, Knysh2013Estimation, Matsumoto2002A, Toth2012Multipartite,Hyllus2012Fisher,Baumgratz2016Quantum,Conlon2023Approaching, Szczykulska2016Multi-parameter, Marzolino2013Precision, Marzolino2015Erratum, Humphreys2013Quantum, Skotiniotis2015Quantum, Knott2016Local,Pezze2017Optimal,Ciampini2016Quantum,Albarelli2019Evaluating,DemkowiczDobrzanski2020Multi-parameter,Hou2020Minimal,Ho2020Multiparameter}. 
Estimating different components of a magnetic field represents a paradigmatic multiparameter estimation problem whose complexity arises from the non-commutativity of spin operators associated with each component \cite{Toth2012Multipartite,Hyllus2012Fisher,Baumgratz2016Quantum,Conlon2023Approaching}.
Even in scenarios where the Hamiltonians corresponding to different parameters commute, the problem can be highly nontrivial and very relevant for practical applications. 
This is the case, for instance, when measuring the gradient and the homogeneous background contribution of a field component.

Magnetometry of this type  can be realized with differential interferometry with two particle ensembles, which has raised a lot of attention in quantum metrology \cite{Landini2014Phase-noise, Eckert2006Differential,
Stockton2007Bayesian, Durfee2006Long-Term,Cable2010Parameter,Snadden1998Measurement, Fixler2007Atom, Altenburg2016Optimized}.
Another possibility is considering spin chains, which can be relevant in trapped cold ions or optical lattices of cold atoms, where we have individual access to the particles \cite{Urizar-Lanz2013Macroscopic, Zhang2014Fitting,
Ng2014Quantum}. 
With magnetic field imaging techniques, one can obtain the field strength at various locations, from which, in turn, the field gradient can be deduced \cite{Wildermuth2006Sensing,Vengalattore2007High,Koschorreck2011High,Zhou2010Precisely}. It is also possible to obtain higher order coefficients of the Taylor expansion of the magnetic field in distributed multi-parameter estimation \cite{Wolk2020Noisy,Hamann2022Approximate,Hamann2024Optimal,bate2025experimentaldistributedquantumsensing}.

Gradient metrology has been considered for ensembles of cold atoms \cite{Apellaniz2018Precision,Altenburg2017Estimation,Urizar-Lanz2013Macroscopic}.
However, an extension of the model to Bose-Einstein condensates (BEC) of spin-$j$ atoms showed that estimating the gradient efficiently is not possible in such systems \cite{Apellaniz2018Precision}.
Even if the BEC is trapped in a double-well potential, the precision of the parameter estimation scales as $(\Delta \theta)^2\sim 1/N$, which is the shot-noise scaling. 
This is true even if the quantum state is highly entangled, for instance, a multiqubit Dicke state or a GHZ state \cite{Apellaniz2018Precision}.

In this paper, we show that with simple local unitary operations, one can obtain quantum states that can provide a $(\Delta \theta)^2\sim 1/N^2$ scaling for the precision of parameter estimation, which means reaching the Heisenberg scaling.
For the Dicke state and  the GHZ state these local operations are phase-flips and bit-flips, respectively, for half of the particles.
This way, the Dicke state, useful for homogeneous magnetic field estimation, can be turned into a state that is useful for gradient magnetometry.

Based on these ideas, we will consider measuring the homogeneous field in the $x$, $y$ or $z$ directions as well as measuring the gradient in those directions.
We present relations that bound the achievable maximal precision in these estimation tasks. These extend the relations found in Refs.~\cite{Toth2012Multipartite,Hyllus2012Fisher} that describe the trade-off when measuring the field in the $x,$ $y,$ and $z$ directions, using the quantum Fisher information.  Since we consider an ensemble of particles divided into two groups, our work is also relevant in recent efforts for quantum information processing with spatially separated particle ensembles in a BEC \cite{Fadel2018Spatial,Kunkel2018Spatially,Lange2018Entanglement}.

Our paper is organized as follows.
In Sec.~\ref{sec:background}, we present the theoretical background needed to describe the system and its metrological usefulness.
In Sec.~\ref{sec:impossibility}, we review recent findings that show that an efficient gradiometry with BECs is not possible.
In Sec.~\ref{sec:localop}, we show that efficient gradiometry is possible if we apply some local operations on the state before the dynamics.
In Sec.~\ref{sec:measurement}, we discuss how to infer the gradient from measurements on the quantum state.

\section{Theoretical background}
\label{sec:background}

We briefly explain the separability concept of quantum states, and we describe the Hamiltonian acting on the system.
We also present the main tool used to compute the achievable precision, the multiparametric Cramér-Rao bound and its properties.

\subsection{Separability of quantum states}

A quantum state of $N$ particles is fully separable if it can be written as a statistical mixture of product states \cite{Werner1989Quantum,Horodecki2009Quantum,Guhne2009Entanglement,Friis2019}
\begin{equation}
    \varrho =\sum_{k} p_k \varrho_{k,1}\otimes \cdots \otimes \varrho_{k,N}.
\end{equation}
Otherwise, the quantum state is entangled.

Let us now divide the particles into two groups, `a' and `b'. A state is bipartite separable if it can be written as
\begin{equation}
    \varrho = \sum_{k} p_k \varrho_{k,\rma}\otimes \varrho_{k,\rmb},
\end{equation}
where $\varrho_{k,\rma}$ and $\varrho_{k,\rmb}$ are quantum states of the two different parties with $N_\rma$ and $N_\rmb$ particles, respectively. 

Whereas all fully separable states are bipartite separable, the opposite statement is not always true.
Bipartite separable states might be entangled within the parties, thus they are not necessarily fully separable.

\subsection{Cramér-Rao bound for multiparameter quantum metrology}
\label{sec:CR_multi}

First, let us summarize briefly basic facts about single-parameter metrology.
In the most fundamental task in quantum metrology, we have to consider a density matrix undergoing the dynamics
\begin{equation}
    \varrho_\theta=U_\theta\varrho U_\theta^\dagger,
\end{equation}
where the unitary is given as 
\begin{equation}
U_\theta=e^{-iA\theta}.
\end{equation}

Based on measurements on $\varrho_\theta,$ we would like to estimate the parameter $\theta$.
The precision of estimating $\theta$ with an unbiased estimator is bounded from below by the Cram\'er-Rao bound as \cite{Helstrom1976Quantum,Holevo1982Probabilistic,Braunstein1994Statistical, Petz2008Quantum,Braunstein1996Generalized,Giovannetti2004Quantum-Enhanced,Demkowicz-Dobrzanski2014Quantum,Pezze2014Quantum,Toth2014Quantum,Pezze2018Quantum,Paris2009QUANTUM}
\begin{equation}
(\Delta \theta)^2\ge\frac1{\nu F[\varrho,A]}.\label{eq:crbound}
\end{equation}
Here, the quantum Fisher information (QFI) is computed as
\cite{Helstrom1976Quantum,Holevo1982Probabilistic,Braunstein1994Statistical,
Petz2008Quantum,Braunstein1996Generalized}
\begin{equation}
    \qfi[\varrho, A] = 2\sum_{\lambda\neq\mu} \frac{(p_\lambda - p_\mu)^2}{p_\lambda + p_\mu} |\braoket{\lambda}{A}{\mu}|^2 ,
\end{equation}
where the density matrix is given by its eigendecomposition as
\begin{equation}
\varrho = \sum_\lambda p_\lambda \projector{\lambda}.
\end{equation}
Moreover, in Eq.~\eqref{eq:crbound} $\nu$ is the number of independent repetitions. The Cramér-Rao bound in Eq.~\eqref{eq:crbound} can be saturated, under certain reasonable assumptions, in the large $\nu$ limit by an optimal measurement and a suitable estimator. 
In the rest of the paper, we will consider $\nu=1$ for simplifying our formulas, however, the results for $\nu>1$ can straightforwardly be obtained.

For two operators, the QFI is defined as 
\begin{equation}
    \label{eq:qfi-extended}
    \qfi[\varrho, A, B] = 2\sum_{\lambda\neq\mu} \frac{(p_\lambda - p_\mu)^2}{p_\lambda + p_\mu} \braoket{\lambda}{A}{\mu} \braoket{\mu}{B}{\lambda},
\end{equation}
which is linear in the second and third arguments, it is also invariant under the permutation of $A$ and $B.$ We can recover the usual QFI by imposing $A = B$. For pure states, ${\hatPsi}:=\projector{\Psi}$, $\qfi[\Psi, A, B] = 4(\expect{\{A, B\}/2}_{\hatPsi} - \expect{A}_{\hatPsi}\expect{B}_{\hatPsi} )$, and hence $\qfi[\Psi, A] = 4\variance{A}_{\hatPsi},$ that is, the QFI equals the variance times four.

In the multiparameter case, the metrological performance of the quantum state is given by the Cram\'er-Rao matrix inequality \cite{Paris2009QUANTUM,Braunstein1994Statistical,Holevo1982Probabilistic,Helstrom1976Quantum,Petz2002Covariance,Petz2008Quantum}
\begin{equation}
\bm{C} \geqslant \qfim^{-1},
\end{equation}
where the elements of the QFI matrix are given as 
\begin{equation}
\qfim_{ij}= \qfi[\varrho, A_i, A_j],
\end{equation}
and the covariance matrix is defined as 
\begin{equation}
    \bm{C}_{ij}={\rm Cov}[\theta_i,\theta_j]=\langle \theta_i\theta_j\rangle-\langle \theta_i\rangle\langle\theta_j\rangle.
\end{equation}
Here $\theta_i$ are the parameters and $A_i$ are the corresponding Hamiltonian generators.
Note that the diagonal elements of $\bm{C}$ are the variances $\variance{\theta_i}$ of the unknown parameters.
 
\subsection{Hamiltonian}

For differential magnetometry, one has to take into account the effects of an unknown homogeneous field.
For simplicity, we restrict the magnetic field to point towards a single $l$-direction, where $l=x,y,z.$
Hence, for small distances in a certain $x$-direction, the magnetic field can be described as a linear function of $x$,
\begin{equation}
    B_l = B_0 + B_1x.
\end{equation}

The simplest one-particle Hamiltonian for the field-particle interaction is
\begin{equation}
    h = \gamma(B_0 \matrixid \otimes j_l + B_1 \hatx \otimes j_l),
\end{equation}
where $j_l$ and $\hatx$ are the $l$-component of the angular momentum and the position operator respectively, and $\gamma=g\mu_{\rm B}$ is the gyromagnetic ratio which encodes the $g$-factor and the Bohr magneton.
The homogeneous field rotates the system globally, whereas the gradient field acts differently depending on the position of the particle.

The Hamiltonian of the whole system is simply the sum of each of the one-particle Hamiltonians
\begin{equation}
    \label{eq:generic-hamiltonian}
    H = \gamma\left(B_0\matrixid \otimes J_l + B_1 \sum_{n=1}^N \hatx_n \otimes j_{l,n} \right),
\end{equation}
where $J_l$ is the $l$-component of the total angular momentum, with $l=x,y,z,$ and $N$ is the total particle number.

The system is split in two ensembles along the $x$-direction, and we place the origin such that the first ensemble is at $x=-d$ and second at $x=+d.$
We can reformulate the Hamiltonian, acting on the spin-subspace only, as
\begin{equation}
    \label{eq:double-well-hamiltonian}
    H = \gamma \left[B_0 J_l + B_1 d (J_{l,\rma} - J_{l,\rmb})\right],
\end{equation}
where `a' stands for particles on the first well and `b' for particles on the second well.
In general, the splitting might not be even and we use $N_\rma$ to refer to the number of particles in the well `a' and $N_\rmb$ for the well `b', where $N=N_\rma +N_\rmb$.
Note also that $J_l = J_{l,\rma} + J_{l,\rmb}$.

The state evolves under the unitary operator $U=\exp(-iH t)$, where $t$ is the evolution time.
The homogeneous phase-shift and the gradient phase-shift imprinted on the state are
\begin{subequations}\label{eq:unknown-parameters}
\begin{align}
    b_0 & = \gamma B_0 t, \\
    b_1 & = \gamma (B_1 d) t,
\end{align}
\end{subequations}
respectively.
Note that the unknown parameter $b_1$ encodes the distance $d$.
Finally, the unitary time-evolution operator takes the form
\begin{equation}
    \label{eq:complete_hamiltonian}
    U = \exp\!\left\{-i\left[b_0 (J_{l,\rma} {+} J_{l,\rmb}) + b_1 (J_{l,\rma} {-} J_{l,\rmb})\right]\right\}\!,
\end{equation}
where we identify $J_{l,\rma} {+} J_{l,\rmb}$ as the homogeneous phase-shift generator and $J_{l,\rma} {-} J_{l,\rmb}$ as the gradient phase-shift generator.

The generators for the homogeneous and the gradient phase-shifts, $J_{l,\rma} {+} J_{l,\rmb}$ and $J_{l,\rma} {-} J_{l,\rmb}$, are related to each other by a ``partial flipping'' operation
\begin{equation}
    \label{eq:def-flipping-op}
    R_{\perp,\rmb}(\pi):= \matrixid_\rma\otimes\exp(-i \pi J_{\perp,\rmb}),
\end{equation}
which rotates by $\pi$ the spin of the particles in `b' around the $\perp$-axis perpendicular to $l$.
Thus, 
\begin{equation}
    R_{\perp,\rmb}(\pi)^\dagger
    (J_{l,\rma} {+} J_{l,\rmb})
    R_{\perp,\rmb}(\pi) = J_{l,\rma} {-} J_{l,\rmb}.
\end{equation}
For instance, we can have $l=z$ and $\perp=x.$

\subsection{Gradient metrology in a two-well system}

Let us now apply the formalism of quantum metrology presented in Sec.~\ref{sec:CR_multi} to the problem considered in our article.
For states insensitive to the gradient of the field, we have that $[\varrho, J_{l,\rma}{-}J_{l,\rmb}] = 0$.
Thus, the only parameter that can be estimated is $b_0$.
Its achievable precision is calculated utilizing the Cramér-Rao bound as 
\begin{equation}
    \label{eq:homo-bound-ins}
    \ivariance{b_0} \leqs \qfi[\varrho, J_{l,\rma} {+} J_{l,\rmb}],
\end{equation}
where the $\qfi$ is the quantum Fisher Information (QFI).

On the other hand, for states insensitive to the homogeneous field, $[\varrho, J_l] = 0$ holds. Thus, the achievable precision for $b_1$ is
\begin{equation}
    \label{eq:grad-bound-ins}
    \ivariance{b_1} \leqslant \qfi[\varrho, J_{l,\rma}{-}J_{l,\rmb}].
\end{equation}
From now on, we will use $\qfii{l,+}:=\qfi[\varrho, J_{l,\rma} {+} J_{l,\rmb}]$ and $\qfii{l,-}:=\qfi[\varrho, J_{l,\rma} {-} J_{l,\rmb}]$ for simplicity.

For states sensitive to both $b_0$ and $b_1$, instead of Eqs.~\eqref{eq:homo-bound-ins} and \eqref{eq:grad-bound-ins}, we have a $2 \times 2$ Cramér-Rao matrix inequality
\begin{equation}\label{eq:cr-matrix-inequality}
    \begin{pmatrix}
        \variance{b_0} & \Cov[b_0,b_1]\\
        \Cov[b_0,b_1] & \variance{b_1}
    \end{pmatrix}\geqslant \qfim_l^{-1},
\end{equation}
where $\Cov[b_0,b_1]$ is the covariance between $b_0$ and $b_1$ given as
\begin{equation}
 \Cov[A,B] = \frac{1}{2}\expect{\{A,B\}} - \expect{A}\expect{B}. \label{eq:Cov}
\end{equation}
The QFI matrix elements are given by
\begin{subequations}\label{eq:FQmat}
    \begin{align}
            (\qfim_l)_{0,0} & = \qfii{l,+} = \qfii{l,\rma} + 2\qfii{l,\rma\rmb} + \qfii{l,\rmb} ,
        \\
        \label{eq:FQmat01}
        \begin{split}
            (\qfim_l)_{0,1} & = (\qfim_l)_{1,0} \\
            & = \qfii{l,\rma} + \qfii{l,\rma\rmb} + \qfii{l,\rma\rmb} - \qfii{l,\rmb} \\
            & = \qfii{l,\rma} - \qfii{l,\rmb},
        \end{split} \\
            (\qfim_l)_{1,1} & = \qfii{l,-}  = \qfii{l,\rma} - 2\qfii{l,\rma\rmb} + \qfii{l,\rmb},
    \end{align}
\end{subequations}
where we use $\qfii{l,rr'}$ for $\qfi[\varrho, J_{l,r},J_{l,r'}]$, $\qfii{l,r}$ for $\qfi[\varrho, J_{l,r}]$
for readability, and the properties of QFI \eqref{eq:qfi-extended} to rewrite and simplify the matrix elements.

Hence, using the algebraic formula for the inverse of a $2 \times 2$ matrix and from the diagonal terms in Eq.~\eqref{eq:cr-matrix-inequality}, the corresponding precision bounds for $b_0$ and $b_1$ are
\begin{subequations}
    \label{eq:bounds}
    \begin{align}
        \label{eq:homo-bound}
        \ivariance{b_0} &\leqs (\qfim_l)_{0,0} - \frac{(\qfim_l)_{0,1}^2}{(\qfim_l)_{1,1}} = 4\frac{\qfii{l,\rma}\qfii{l,\rmb} - \qfii{l,\rma\rmb}^2}{\qfii{l,-}}, \\
        \label{eq:grad-bound}
        \ivariance{b_1} &\leqs (\qfim_l)_{1,1} - \frac{(\qfim_l)_{0,1}^2}{(\qfim_l)_{0,0}} = 4\frac{\qfii{l,\rma}\qfii{l,\rmb} - \qfii{l,\rma\rmb}^2}{\qfii{l,+}}.
    \end{align}
\end{subequations}
Since the two generators commute with each other, it is possible to saturate both bounds at the same time \cite{Ragy2016Compatibility,Liu2020QFIMatrix}.
Note that for homogeneous and gradient precision, we have in the denominator the QFI that corresponds to the gradient and the homogeneous parameter, respectively.
It suggests that the better the state for estimating the homogeneous field, the worse for estimating the gradient of the field, and \textit{vice-versa}.

Moreover, if $\qfi[\varrho, J_{l,\rma}] = \qfi[\varrho, J_{l, \rmb}]$, then the off-diagonal elements of the QFI matrix, $(\qfim_l)_{0,1}$ and $(\qfim_l)_{1,0}$, vanish as can be seen in Eq.~\eqref{eq:FQmat01}.
Hence, the QFI matrix is diagonal, and we recover the precision bounds for $b_0$ and $b_1$ as
\begin{subequations}
    \begin{align}
        \label{eq:homo-bound-sym}
        \ivariance{b_0} &\leqs \qfi[\varrho, J_l], \\ 
        \label{eq:grad-bound-sym}
        \ivariance{b_1} &\leqs \qfi[\varrho, J_{l,\rma}{-}J_{l, \rmb}]. 
    \end{align}
\end{subequations}
The bounds are the same as if the state was insensitive to the corresponding second parameter, Eqs.~\eqref{eq:homo-bound-ins} and~\eqref{eq:grad-bound-ins}, \ie a single-parameter estimation problem.

It is instructive to consider the extreme cases when either $N_\rma$ or $N_\rmb$ is zero.
In this case, all the particles are in one of the wells and therefore, cannot be used for gradient magnetometry.
If $N_\rma=0$ then the Hamiltonian is simplified from Eq.~\eqref{eq:complete_hamiltonian} to
$H = (b_0 - b_1) J_l$, and one can only estimate the value of the field at wells `b', and estimate $b_0 - b_1$.
On the other hand, if $N_\rmb=0$ then the Hamiltonian simplifies from Eq.~\eqref{eq:complete_hamiltonian} to
$H = (b_0 + b_1) J_l$, and one can only estimate the value of the field at well `a', and estimate $b_0 + b_1$, see Ref.~\cite{Fadel2022Multiparameter}.

We now analyze the joint performance for estimating both parameters. 
Using the properties of the QFI given in Eq.~\eqref{eq:qfi-extended}, we obtain the sum of the bounds in Eqs.~\eqref{eq:bounds} as
\begin{equation}
    \begin{split}
        \ivariance{b_0}+\ivariance{b_1}\leqs \left(\qfii{l,+} + \qfii{l,-}\right)\left[1-\frac{\big(\qfii{l,\rma}- \qfii{l,\rmb}\big)^2}{\qfii{l,+}\qfii{l,-}}\right].
    \end{split}
\end{equation}
Note that the term in the square bracket on the right-hand side of the inequality is equal to or smaller than one.
Hence, we can bound the sum of the achievable precisions by
\begin{equation}
    \label{eq:sum-bounded-by-qfi}
    \begin{split}
        \ivariance{b_0}+\ivariance{b_1} & \leqs \qfii{l,+} + \qfii{l,-} = 2\left(\qfii{l,\rma} + \qfii{l,\rmb}\right).
    \end{split}
\end{equation}
We obtain the same inequality, if we add the inequalities for single parameter estimation tasks given in Eqs.~\eqref{eq:homo-bound-ins} and \eqref{eq:grad-bound-ins}.

For fully separable states, the right-hand side of Eq.~\eqref{eq:sum-bounded-by-qfi} is tightly bounded from above by $2(N_{\rm{a}} + N_{\rm{b}}) = 2N$. 
Which means that there are separable states for which
\begin{equation}
    \begin{split}
\qfi[\varrho, J_{l, \rm{a}}{+} J_{l, \rm{b}}]& = N,\\
\qfi[\varrho, J_{l, \rm{a}}{-} J_{l, \rm{b}}]& = N,
    \end{split}\label{eq:twoineq}
\end{equation}
hold if $\qfii{l,\rma} = \qfii{l,\rmb}$.

On the other hand, for entangled states, we have that
\begin{equation}
    \label{eq:Jplusminus}
    \qfii{l,+} + \qfii{l,-}\leqs 2(N_\rma^2+N_\rmb^2)\leqs N^2,
\end{equation}
where the second inequality is saturated for an evenly split state, for which $N_\rma=N_\rmb.$
The inequality in Eq.~\eqref{eq:Jplusminus} tells us that the Heisenberg limit, $\ivariance{\theta} \leqs N^2$, cannot be saturated for estimating the homogeneous field and the gradient field at the same time.
Note also that the  bound in Eq.~\eqref{eq:Jplusminus} can be saturated if the state is of the form $\ket{\Psi_\rma}\otimes\ket{\Psi_\rmb},$ which is a tensor product of two single-subsystem states.
Nevertheless, with such states one can only reach half of the precision limit $\qfii{l,\pm} = N^2/2$ for each of the parameters, since $\qfii{l,\rma\rmb} = 0.$

\section{Impossibility of efficient gradient metrology with BECs}
\label{sec:impossibility}

We briefly review the findings of Ref.~\cite{Apellaniz2018Precision} concerning gradiometry with a single BEC. We show that the precision of gradient metrology cannot surpass the shot-noise limit.

In a BEC all the particles share the same spatial state $\projector{\Psi}$.
Hence, the whole state can be written as $\projector{\Psi}^{\otimes N} \otimes \varrho$, where $\varrho$ is the spin state.
The Hamiltonian is of the form Eq.~\eqref{eq:generic-hamiltonian}, and
\begin{subequations}
\begin{align}
    \mu & = \braoket{\Psi}{\hatx_n}{\Psi}, \\
    \sigma^2 & = \braoket{\Psi}{(\hatx_n - \mu)^2}{\Psi} = \braoket{\Psi}{\hatx_n^2}{\Psi} - \mu^2
\end{align}
\end{subequations}
are the mean position and the variance of the position of the BEC, respectively.
The corresponding multiparameter phase-shift generators are $J_l = \sum_{n=1}^N j_{l,n}$ and $\sum_{n=1}^N \hatx_n \otimes j_{l,n}$. Note that in this section the generators are different from the case considered in Eq.~\eqref{eq:FQmat}, where the generators were simply $J_{l,\rma} {\pm} J_{l,\rmb}.$
We compare the two systems at the end of the section.

The QFI matrix elements for pure states are
\begin{subequations}
    \begin{align}
        (\qfim_l)_{0,0} & = 4 \variance{J_l}, \\
        \begin{split}
        (\qfim_l)_{0,1} & = (\qfim_l)_{1,0} \\
        & = 4\left(\expect{J_l \sum_{n=1}^N \hatx_n \otimes j_{l,n}} - \expect{J_l}\expect{\sum_{n=1}^N \hatx_n \otimes j_{l,n}}\right) \\ & = 4\mu \variance{J_l} = 0,
        \end{split}\\
            (\qfim_l)_{1,1} & = 4\left( \expect{\biggl(\sum_{n=1}^{N}\hatx_n \otimes j_{l,n}\biggr)^2}
            - \expect{\sum_{n=1}^N \hatx_n \otimes j_{l,n}}^2\right) \nonumber\\
            & = 4\left( \sigma^2 \expect{\sum_{n=1}^{N}j_{l,n}^2}
            + \mu^2 \expect{\sum_{n, m = 1}^{N}j_{l,n} j_{l,m}} - \mu^2\expect{J_l}^2\right) \nonumber \\
            & = 4\left( \sigma^2 \expect{\sum_{n=1}^{N}j_{l,n}^2}
            + \mu^2 \variance{J_l}\right) \\
            & = 4\sigma^2 \expect{\sum_{n=1}^{N}j_{l,n}^2}, \nonumber
    \end{align}
\end{subequations}
where we set $\mu=0$ without loss of generality. 
Hence, the achievable precision is bounded from above as
\begin{equation}
    \ivariance{b_1} \leqs (\qfim_l)_{1,1} - \frac{(\qfim_l)_{0,1}^2}{(\qfim_l)_{0,0}}  \leqs 4\sigma^2 \expect{\sum_{n=1}^{N}j_{l,n}^2}\leqs\sigma^2 N,
\end{equation}
which for the last inequality we considered $N$ spin-$\frac{1}{2}$ particles.  We can already see the linear dependence on $N$ corresponding to the shot-noise scaling.

Let us now make a quantitative comparison to the results of this paper.
Let us consider  $\sigma^2$ appearing in the the upper-bound. 
We compute $\sigma^2$ for the double well.
For systems described with Eqs.~\eqref{eq:double-well-hamiltonian}, the size of the system $\sigma^2$ is given by
\begin{equation}
    \sigma^2 = \frac{\sum_{n=1}^N d^2}{N} - \left(\frac{\sum_{n=1}^{N_\rma} (-d) + \sum_{n=1}^{N_\rmb} d}{N}\right)^2 = d^2\frac{4N_\rma N_\rmb}{N^2}.
\end{equation}
The parameter $d$ is absorbed into $b_1$, see Eq.~\eqref{eq:unknown-parameters}.
Hence, the achievable precision for the BECs can be written as 
\begin{equation}
    \ivariance{b_1}_{\rm BEC} \leqs \frac{4N_\rma N_\rmb}{N}\label{eq:BEC}
\end{equation}
where $N_\rma$ and $N_\rmb$ belong to the double-well system the BEC is compared against.
If we have an evenly split condensate then the bound in Eq.~(\ref{eq:BEC}) becomes
\begin{equation}
  \ivariance{b_1}_{\rm BEC} \leqs N,
\end{equation}
which is just the shot-noise limit.

\section{Flipping operator and partially flipped Dicke state for gradiometry}
\label{sec:localop}

Efficient gradiometry is still possible if we carry out simple local operations on the quantum state before the gradient measurement.
In particular, based on Eq.~\eqref{eq:def-flipping-op} and Eqs.~\eqref{eq:homo-bound-sym} and~\eqref{eq:grad-bound-sym}, one can apply the unitary flipping operator $R_{\perp, \rmb}(\pi)$ over the states that have high sensitivity on the homogeneous field to obtain states that have high sensitivity on estimating the gradient of the field.
The opposite is also true.

To illustrate this, one can take for example the Greenberger–Horne–Zeilinger (GHZ) state with half of the particles at each well to estimate magnetic fields pointing towards the $z$-direction.
The GHZ state
\begin{equation}
    \label{eq:ghz}
    \ket{{\rm GHZ}} = \frac{1}{\sqrt{2}} (\ket{0}^{\otimes N}
    + \ket{1}^{\otimes N})
\end{equation}
commutes with $J_{z,\rma}{-}J_{z,\rmb}$.
Hence, the corresponding bound for $b_0$ is
\begin{equation}
    \ivariance{b_0} \leqs 4 \variance{J_z}_{\rm GHZ} = N^2.
\end{equation}
If we apply the flipping operator
\begin{equation}
    \label{eq:rotation-x}
    R_{x,\rmb}(\pi) = \mathds{1}\otimes \sigma_x^{\otimes N/2}
\end{equation}
to the GHZ state along the $x$-direction, we obtain
\begin{equation}
    \label{eq:ghz-flipped}
    \ket{\widetilde{\rm GHZ}} = \frac{1}{\sqrt{2}}
    (\ket{0}_{\rma\phantom{\rmb}}^{\otimes N/2}\ket{1}_{\rmb}^{\otimes N/2}
    +\ket{1}_{\rma\phantom{\rmb}}^{\otimes N/2}\ket{0}_{\rmb}^{\otimes N/2}),
\end{equation}
which commutes with $J_z$.

The corresponding bound for the estimation of $b_1$ is optimal,
\begin{equation}
    \ivariance{b_1} \leqs 4 [\Delta (J_{z,\rma}{-}J_{z,\rmb})]^2_{\widetilde{\rm GHZ}} = N^2.
\end{equation}
In the following we take this approach to construct states that are sensitive to gradient fields based on states that are sensitive to homogeneous fields \cite{Fadel2022Multiparameter}.

\subsection{Partially flipped Dicke states}

The Dicke state can be used to measure the homogeneous magnetic field.
In this section, we will show, how to obtain a quantum state from the Dicke state via local unitaries, that is useful for gradient metrology.

In general, Dicke states are simultaneous eigenstates of $ \bm{J}^2$ and $J_z$
\begin{align*}
    \bm{J}^2\ket{j,m_z}&= j(j+1)\ket{j,m_z},\\
    J_z\ket{j,m_z}&=m_z\ket{j,m_z}.
\end{align*}
We are mostly interested in the Dicke state of an $N$-qubit system for which $j=N/2$ and $m_z=0$,
\begin{equation}
    \ket{{\rm D}_N} \equiv \ket{N/2, 0}.
\end{equation}
The $\ket{{\rm D}_N}$ state is a symmetric
with $\expect{J_z}=0$ and $\variance{J_z}=0,$ and large variances $\variance{J_x}$ and $\variance{J_y}$.
Since it is a pure state, we have
\begin{equation}
\qfi[\varrho, J_z]=4\variance{J_z}=0.
\end{equation}
Hence, the Dicke state is insensitive to rotations around the $z$-axis.
On the other hand, for this state,
\begin{equation}
    \qfi[\varrho, J_l]=4\variance{J_l}=\frac{N(N+2)}{2},
\end{equation}
for $l=x,y$, is large, thus the state can be used for phase estimation over rotations around the $x$- and $y$- axes, or any axis in the $xy$-plane.

For spin-$\frac{1}{2}$ particles, the Dicke state $\ket{{\rm D}_N}$ can be written as
\begin{equation}
    \label{eq:dicke}
    \ket{{\rm D}_N} =  \binom{N}{N/2}^{-1/2}
    \sum_{k} \mathcal{P}_k(\ket{0}^{\otimes N/2}\ket{1}^{\otimes N/2}),
\end{equation}
where the summation is over all distinct permutations of the spins.

A Schmidt decomposition of the Dicke state given in Eq.~(\ref{eq:dicke}) corresponding a bipartitioning to $N_\rma$ and $N_\rmb$ spins can be written as \cite{Stockton2003Characterizing}
\begin{equation}
    \label{eq:dicke-subhilb}
    \ket{{\rm D}_N} = \sum_{m=-N_\rma/2}^{N_\rma/2}
        c_m
        \ket{m}_\rma\ket{{-}m}_\rmb,
\end{equation}
where we define the expression 
\begin{equation}
    c_m={\binom{N}{N/2}^{-1/2}} 
        \sqrt{{\binom{N_\rma}{N_\rma/2+m}}
        {\binom{N_\rmb}{N_\rmb/2+m}}},\label{eq:cm}
\end{equation}
and the coefficients appearing in Eq.~\eqref{eq:cm} are usual Clebsch–Gordan coefficients.
We consider $N_\rma \leqs N_\rmb$ without loss of generality, and we used the notation  $\ket{m}_\rma=\ket{N_{\rma}/2, m}_\rma$ and $\ket{m}_\rmb=\ket{N_\rmb / 2, m}_\rmb.$

Since the Dicke state is very sensitive to rotations around the $x$- and $y$-axis, we will apply the flipping operator $R_{z,\rmb}(\pi)$ along the $z$-axis,
\begin{equation}
    R_{z,\rmb}(\pi) = \mathds{1}\otimes \sigma_z^{\otimes N/2}.
\end{equation}
Hence, the partially flipped Dicke state, $\ket{\widetilde{\rm D}_N}:=R_{z,\rmb}(\pi)\ket{{\rm D}_N}$, can be used for gradient magnetometry along any direction perpendicular to $z$-axis.
$R_{z,\rmb}(\pi)$ acts only on $\ket{m}_\rmb$, and based on Eqs.~\eqref{eq:def-flipping-op} and \eqref{eq:dicke-subhilb}, we have
\begin{equation}
    \label{eq:dicke-twist}
    \ket{\widetilde{\rm D}_N} =\sum_{m=-N_\rma/2}^{N_\rma/2}
    c_m
    (-1)^{m}
    \ket{m}_\rma\ket{{-}m}_\rmb.
\end{equation}

\subsection{Gradient magnetometry with partially flipped Dicke states}
\label{ssec:gradient-with-f-dicke-polytope}

For the task of estimating the three components of a magnetic field a quantum state cannot surpass the following precision bound \cite{Toth2012Multipartite,Hyllus2012Fisher}
\begin{equation}
    \label{eq:fig-merit-hom}
    \sum_{l= x,y,z} \qfi[\varrho, J_l] \leqs 4 \lambda_{\max} (\bm{J}^2) =
     N(N + 2),
\end{equation}
where $\lambda_{\max} (X)$ is the maximal eigenvalue of the matrix $X$.
The GHZ and Dicke states among others saturate this bound.

\begin{observation}
    \label{obs:lmax-var-diff}
    For differential magnetometry a similar figure of merit would be
    \begin{multline}
        \label{eq:fig-merit-gra}
        \sum_{l= x,y,z} \qfi[\varrho, J_{l,\rma}{-}J_{l,\rmb}] \leqs
         N(N + 2) + 4\min(N_\rma, N_\rmb),
    \end{multline}
    which tells us how well a state $\varrho$ can estimate the gradient parameters over all directions. Note that the sum of the quantum Fisher information terms corresponding to gradient measurements in the three orthogonal directions in Eq.~\eqref{eq:fig-merit-gra} can be larger than the upper bound on the sum of the quantum Fisher information terms corresponding to measurements of the homogeneous magnetic field in the three directions given in Eq.~\eqref{eq:fig-merit-hom}. For a proof, see Appendix~\ref{app:proof-fig-merit-gra}.
\end{observation}

The states that maximize the left-hand side of Eq.~\eqref{eq:fig-merit-gra} belong to the subspace in which the total spin number is minimal, $j=|N_\rma{-}N_\rmb|/2$. For an evenly split state, the total angular momentum number is zero which corresponds to a bipartite singlet state, see Appendix~\ref{app:proof-fig-merit-gra}.

It is easy to see that for the flipped GHZ the left-hand side of the Eq.~\eqref{eq:fig-merit-gra} gives us $N(N+2)$.
Whereas for the flipped Dicke state, the expression on the left-hand side of the Eq.~\eqref{eq:fig-merit-gra} is
\begin{equation}
    \label{eq:f-dicke-better}
    \sum_{l= x,y,z} \qfi[\varrho, J_{l,\rma}{-}J_{l,\rmb}] = N(N+2)+\frac{4N_\rma N_\rmb}{N-1}.
\end{equation}
The flipped Dicke is closer to saturate the Eq.~\eqref{eq:fig-merit-gra} than the flipped GHZ state.
See Appendix~\ref{app:expectation-values} for more details on computing some expectation values for the Dicke and flipped Dicke states.

If we flip the ensemble `b' in the $x,y$ or $z$ basis, then the sign of two of the $J_{l,\rmb}$ terms in Eqs.~\eqref{eq:fig-merit-hom} or \eqref{eq:fig-merit-gra} changes, while the bounds remain the same. 
Hence, we can reformulate the entire set of six equations as
\begin{multline}
    \label{eq:planes}
    \sum_{l = x,y,z}\qfi[\varrho, J_{l,\rma}{+}s_lJ_{l,\rmb}] \\ \leqs N(N + 2) + 4\,\Pi(s_xs_ys_z) \min(N_\rma, N_\rmb),
\end{multline}
where $s_l= +1,-1$ for $l=x,y,z,$ and we define the parity function as $\Pi(1)=0$ and $\Pi(-1)=1$.
 
\begin{figure}[htb]
    \centering
    \includegraphics[width=\columnwidth]{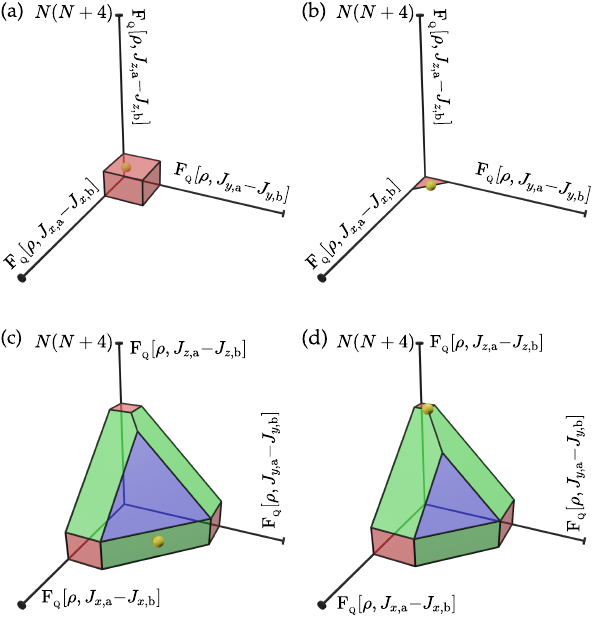}
    \caption{
        Polytopes for different quantum states showing the allowed values of $\qfi[\varrho, J_{l,\rma}{-}J_{l,\rmb}]$ for $l=x,y,z$, for given values of $\qfi[\varrho, J_{l,\rma}{+}J_{l,\rmb}]$ for $l=x,y,z$ and for $N=8$ particles,
        based on the set of inequalities Eq.~\eqref{eq:gradient-planes}.
        (red), (green) and (blue) polygons correspond to planes obtained by Eqs.~\eqref{eq:x-plane}, \eqref{eq:xy-plane} and \eqref{eq:xyz-plane}, respectively.
        (yellow ball) Actual values of $\qfi[\varrho, J_{l,\rma}{-}J_{l,\rmb}]$ for $l=x,y,z$ for the given quantum state.
        (a) Dicke state. 
        The ball is inside the cuboid, indicating that the Dicke state does not saturate any of the bounds.
        (b) GHZ state, saturating the inequality \eqref{eq:xy-plane} for $i,j,k=x,y,z$ respectively. 
        (c) Partially flipped Dicke state, saturating the inequality \eqref{eq:xy-plane} for $i,j,k=x,y,z$ respectively.
        (d) Partially flipped GHZ state, saturating the inequalities \eqref{eq:xy-plane} for $i,j,k=y,z,x$ and $i,j,k=z,x,y$, and Eq.~\eqref{eq:x-plane} for the $z$-direction.
        For the partially flipped Dicke state, the ball is closer to the blue plane than in the case of the flipped GHZ state, as suggested by Eq.~\eqref{eq:f-dicke-better} (For details, see text).
        We compute the values obtained in each case in Appendix \ref{app:polytope-expectation-values}.
    }
    \label{fig:polytopes}
\end{figure}

These inequalities in Eq.~\eqref{eq:planes}, together with the first inequality in Eq.~\eqref{eq:Jplusminus}, define a convex polytope which confines the allowed values for $\qfi[\varrho, J_{l,\rma}{\pm}J_{l,\rmb}]$ based on $N_\rma$ and $N_\rmb.$
Moreover, one can find upper-bounds for $\qfi[\varrho, J_{l,\rma}{-}J_{l,\rmb}]$, the ability of the state to sense the differential of the field, based on $\qfi[\varrho, J_{l,\rma}{+}J_{l,\rmb}]$, the ability of the state to sense the homogeneous magnetic field.
We find the following upper-bounds, 
\begin{subequations}
    \label{eq:gradient-planes}
    \begin{align}
        \qfii{i,-} \leqslant \min\!\big(N^2 - \,&\qfii{i,+}, \nonumber\\
        N(N{+}2) + 4 & \min(N_\rma, N_\rmb) - \qfii{j,+} - \qfii{k,+}\big),  \label{eq:x-plane}\\
        \qfii{i,-} + \qfii{j,-} & \leqslant N(N+2) - \qfii{k,+}, \label{eq:xy-plane}\\
        \qfii{i,-} + \qfii{j,-} + \qfii{k,-} & \leqslant  N(N+2) + 4 \min(N_\rma, N_\rmb), \label{eq:xyz-plane}
    \end{align}
\end{subequations}
where $i, j, k$ take all permutations of $x, y, z.$ 
We show this polytope for the case of $N=8$ particles and for different states 
in Figure~\ref{fig:polytopes}.

\subsection{Precision bound for partially flipped Dicke state}

We study the precision bound for the gradient parameter $b_1$, Eq.~\eqref{eq:grad-bound}, for a magnetic field pointing towards the $y$-direction when the initial state is a flipped Dicke state given in Eq.~\eqref{eq:dicke-twist}.
First, we compute the quantities $\qfii{y,\rma}$, $\qfii{y,\rmb}$ and $\qfii{y,\rma,\rmb}$ appearing in Eq.~\eqref{eq:grad-bound}
\begin{subequations}
\begin{align}
    \qfii{y,\rma} & = 4\variance{J_{y,\rma}} = 4\expect{J_{y,\rma}^2} = \frac{N_\rma[(N_\rma+1)N-2]}{2(N-1)}, \\
    \qfii{y,\rmb} & = 4\variance{J_{y,\rmb}} = 4\expect{J_{y,\rmb}^2} = \frac{N_\rmb[(N_\rmb+1)N-2]}{2(N-1)}, \\
    \qfii{y,\rma\rmb} & = 4\Cov[J_{y,\rma},J_{y,\rmb}] = 4\expect{J_{y,\rma}J_{y,\rmb}} =-\frac{N_\rma N_\rmb N}{2(N-1)},
\end{align}
\end{subequations}
where  $\Cov[A,B]$ is given as in Eq.~(\ref{eq:Cov}).
Then, the optimal precision bound for $b_1$ is obtained as
\begin{equation}
    \label{eq:bound-grad-twited-dicke-general}
    \ivariance{b_1}_{\max} = \frac{2N_\rma N_\rmb(N^2-4)}{N[N-2+(N_\rma{-}N_\rmb)^2]}.
\end{equation}
For an evenly split state, we have
\begin{equation}
    \label{eq:bound-grad-twited-dicke}
    \ivariance{b_1}_{\max} = \frac{N(N+2)}{2}
\end{equation}
as expected from Eq.~\eqref{eq:grad-bound-sym}.

\begin{observation}
    For a fixed number of particles $N$, the best metrological performance is attained if the ensemble is split evenly.
    Let us consider states with not evenly split subsystems, keeping in mind that $N$ must be even for the $\ket{\widetilde{{\rm D}}_N}$ state.
    We find that adding particles by pairs to one of the wells decreases the achievable precision \eqref{eq:bound-grad-twited-dicke-general}.
\end{observation}
{\it Proof.} The first statement can be proved, since substituting $N_\rmb$ by $N-N_\rma$ in Eq.~\eqref{eq:bound-grad-twited-dicke-general}, we find its maximum at $N_\rma = N/2$.
For the second statement, we need to
substitute $N_\rma$ by $N/2$, $N_\rmb$ by $N/2+n$, and $N$ by $N+n.$ Hence, we obtain
\begin{equation}
    \ivariance{b_1}_{\max}=\frac{N(N+2n)[(N+n)^2-4]}{2(N+n)(N-2+n+n^2)},
\end{equation}
where $N$ is the original particle number and $n$ are the particles added to one of the wells.
For $n=2$ this function is smaller than Eq.~\eqref{eq:bound-grad-twited-dicke-general} and it monotonically decreases with an increasing $n.$
For $n\gg 1$, we have that $\ivariance{b_1}_{\max} = N$ even though the total particle number is now $N+n$. \hfill$\qed$

Finally, we find that for the product of two Dicke states, the achievable precision is $\ivariance{b_1}_{\max} = N/2(N/2+2),$ which is just twice the achievable precision obtained in one of the wells for sensing the homogeneous field \cite{Apellaniz2018Precision}. It is also two times smaller than achievable precision obtained for a general quantum state given in Eq.~\eqref{eq:bound-grad-twited-dicke} for $N\gg1$.

We have concluded the study of the partially flipped Dicke state split into two parts.
See Ref.~\cite{Fadel2022Multiparameter} for the related problem of multiparameter quantum metrology with an ensemble spit in several subsystems. 

\section{Measurement operators for partially flipped Dicke state}
\label{sec:measurement}

In this section, we will look for an operator $\mathcal M$ to estimate the gradient parameter $b_1$ that is not affected by the unitary dynamics
\begin{equation}
\exp[-i b_0(J_{y,\rma}{+}J_{y,\rmb})]
\end{equation}
generated by the homogeneous field parameter $b_0$, c.~f.  Eq.~\eqref{eq:complete_hamiltonian}.

Thus, we impose the condition
\begin{equation}
   [J_{y,\rma}{+}J_{y,\rmb},\mathcal{M}]=0. \label{eq:commJy}
\end{equation}
Based on the error-propagation formula, the achievable precision is
\begin{equation}
    \label{eq:epf}
    \ivariance{b_1}_{\rm epf} = \frac{|\partial_{b_1}\expect{\mathcal{M}}_\varrho|^2}{\variance{\mathcal{M}}_\varrho}
    = \frac{|\expect{[J_{y,\rma}{-}J_{y,\rmb},\mathcal{M}]}_\varrho|^2}{\variance{\mathcal{M}}_\varrho}.
\end{equation}
Note that it would sufficient to require
\begin{equation}
   \langle [J_{y,\rma}{+}J_{y,\rmb},\mathcal{M}]\rangle=0, \label{eq:commJy2}
\end{equation}
for partially flipped Dicke states. However, even with the more stringent condition in Eq.~\eqref{eq:commJy} we will obtain an optimal operator,
the calculations will be simpler, and our results will be valid for any quantum state.

\subsection{Optimal measurement operator}
The procedure to find an optimal operator is as follows.
Let us define for a given $\varrho$ (in this case the flipped Dicke state) a set of $K$ operators
\begin{equation}
    \label{eq:set-of-operators}
    \{M_k\}_{k=1}^{K}
\end{equation}
that fulfill the conditions
\begin{subequations}
\begin{align}
 [J_{y,\rma}{+}J_{y,\rmb},M_k]&=0,\label{eq:Mkeq1}\\
\variance{M_k}_\varrho&\neq0.
\end{align}
\end{subequations}
We require that the gradient phase-shift generator can be written as $J_{y,\rma}{-}J_{y,\rmb} = \sum_{k=1}^K n_k M_k$,
and the optimal measurement operator as $\mathcal{M}_\mathrm{opt} = \sum_{k=1}^K m_k M_k$, where ${\bf n}$ and ${\bf m}$ are $K$-element real vectors.

Then, based on Ref.~\cite{Gessner2019Optimal-measurement}, the optimal precision is obtained as
\begin{equation}
    \label{eq:opt_epf}
    \ivariance{b_1}_{\rm opt}
    = \bm{n}^\mathsf{T}\bm{C}^\mathsf{T}\bm{\Delta}^{-1}\bm{C}\bm{n},
\end{equation}
and the optimal measurement is characterized by
\begin{equation}
    \bm{m} = \alpha \bm{\Delta}^{-1}\bm{C}\bm{n},
\end{equation}
where $\bm{C}_{k,l}=-i\expect{[M_k,M_l]}_\varrho$ and $\bm{\Delta}_{k,l}=\Cov_\varrho[H_k,H_l]$.

We start by defining the set of $M_k$ operators. Let us consider only the symmetric subspace in the two subsystems and the operators acting in this subspace, as such a basis will be sufficient 
to find the optimal operators.  They live in the space determined by the basis states
\begin{equation}
\ket{{\rm D}_{N_{\rma}}^{(m_\rma)}}_y\otimes\ket{{\rm D}_{N_{\rmb}}^{(m_\rmb)}}_y,
\end{equation}
where $m_{\rma}=0,1,...,N_{\rma},$ $m_{\rmb}=0,1,...,N_{\rmb},$ and the Dicke states of $N$ particles with $m$ excitations is defined as
\begin{equation}
    \label{eq:dicke2}
    \ket{{\rm D}_N^{(m)}} =  \binom{N}{N/2}^{-1/2} 
    \sum_{k} \mathcal{P}_k(\ket{0}^{\otimes (N-m)}\ket{1}^{\otimes m}),
\end{equation}
and the $y$ subscript indicates that the Dicke states are in the $y$-basis, rather than in the $z$-basis. Equivalently, we can map the problem to determine the operators acting on two particles, as discussed in Appendix~\ref{app:proof-fig-merit-gra}. The operators must be given in the 
\begin{equation}
\ket{j_{y,\rma}}_y\otimes\ket{j_{y,\rmb}}_y
\end{equation}
basis, where $\ket{j_{y,\rma}}_y$ are eigenstates of the $J_{y,\rma}$ operator with eigenvalue $j_{y,\rma},$ and they are given in the $y$-basis. The states $\ket{j_{y,\rmb}}_y$ are defined analgously.

Since all operators must commute with $J_y\equiv J_{y,\rma}{+}J_{y,\rmb}$ as expressed by Eq.~\eqref{eq:Mkeq1}, they have a block diagonal structure in the basis we use. 
The dimension of the state we have to consider grows rapidly with the size of the system.
This implies that the number of operators one has to consider scales with $\mathcal{O}(N_{\min}^2 N_{\max} N)$, see Appendix~\ref{app:independent-hermitian-operators}.
On the other hand, we can consider the generator of the gradient phase-shift, $M_0=J_{y,\rma}{-}J_{y,\rmb},$ as an additional operator for the set Eq.~\eqref{eq:set-of-operators}. 
With this choice $\bm{n} = \delta_{0,k}\equiv(1,0,0,...,0)^T,$ in which case, the calculation of Eq.~\eqref{eq:opt_epf} simplifies and only  a single column of the matrix $\bm{C}_{k,l}$ needs to be computed as described in Ref.~\cite{Gessner2019Optimal-measurement}. 
However, the method still works for relatively small systems only.

\begin{figure}[htp]
    \centering
    \includegraphics[width=1\columnwidth]{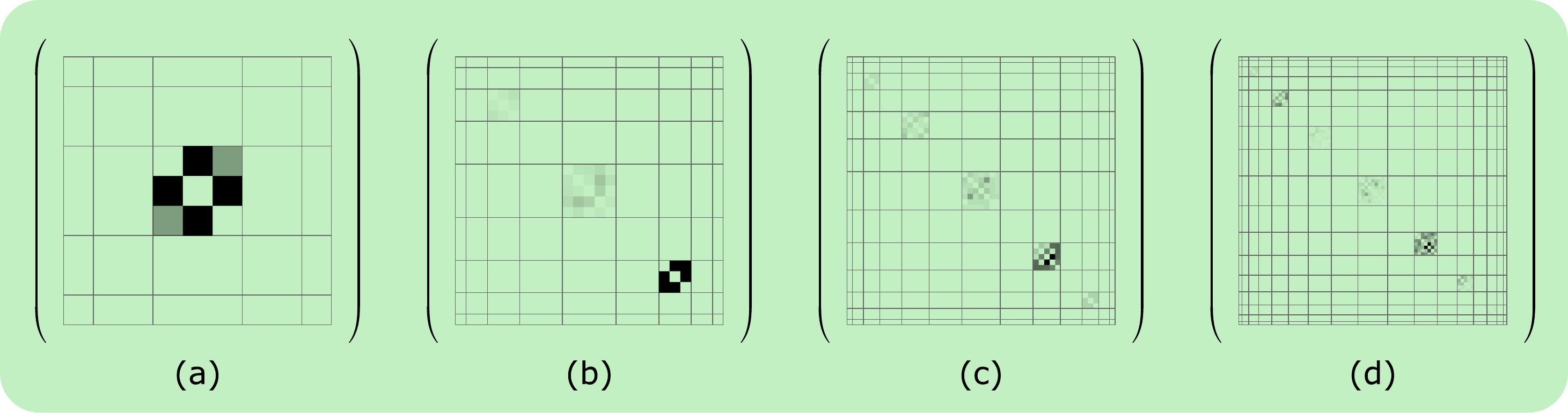}
    \caption{
        (a-d) Reconstruction of the optimal operator $\mathcal{M}_{\rm opt}$ for $N_{\rma}=N_\rmb=2,4,6,8,$ respectively.
        (transparent-boxes) Zero elements. 
        (gray-boxes) Nonzero matrix elements.
        (gray-lines) Boundaries for the different $j_y$ subspaces for the $J_{y,\rma}{+}J_{y,\rmb}$ angular momentum.
        The operator is a block-diagonal operator in the $J_{y,\rma}{+}J_{y,\rmb}$ basis. 
        All the non-zero elements are in the 3rd, 5th, 7th, 9th, etc. diagonal blocks and they are imaginary, while we have zero in the first and the last blocks, which are of size $1\times1.$
    }
    \label{fig:opt-mat}
\end{figure}

Our method makes it possible to obtain explicitly the optimal operator saturating the Cramér-Rao bound~\eqref{eq:bound-grad-twited-dicke-general} for not too large system sizes, see Figure~\ref{fig:opt-mat}.
These results suggest that the Cramér-Rao bound can be saturated with such operators, even for large systems.

\subsection{Measurement operator based on moments of local angular momentum operators}
\label{sec:Measurement_operator_based_on_moments}

We analyze measurement operators as function of the local angular momentum operators $J_{x,\rma}$, $J_{x,\rmb}$, $J_{z,\rma}$ and $J_{z,\rmb}$ in order to make the estimation task more accessible for experiments.
Among all measurements that are insensitive to the homogeneous field, we choose those that maximize the error-propagation formula \eqref{eq:epf} for $b_1$.

We identify one of such operators as
\begin{equation}
    \mathcal{M} = J_{z,\rma}J_{x,\rmb} - J_{x,\rma}J_{z,\rmb}.\label{eq:M2}
\end{equation}

One can compute the derivative of the operator $\mathcal{M}$ with respect to $b_1$ at $t=0$ as
\begin{equation}
    \begin{split}
        |\partial_{b_1}\expect{\mathcal{M}}_\varrho|^2
        & = |\expect{[J_{y,\rma}{-}J_{y,\rmb},\mathcal{M}]}_\varrho|^2 \\
        & = |{-}2\expect{J_{x,\rma}J_{x,\rmb}+J_{z,\rma}J_{z,\rmb}}_\varrho|^2 \\
        & = \left(\frac{N_\rma N_\rmb (N+2)}{4(N-1)}\right)^2,
    \end{split}
\end{equation}
for the partially flipped Dicke state, see Appendix~\ref{app:expectation-values} for more details.
Then, in order to compute the variance $\variance{\mathcal{M}}$, one can calculate first the expectation value of the first moment,
\begin{equation}
    \label{eq:expect-of-m2}
    \expect{\mathcal{M}}_\varrho=\expect{J_{z,\rma}J_{x,\rmb}}_\varrho - \expect{J_{x,\rma}J_{z,\rmb}}_\varrho = 0
\end{equation}
which is zero, due to that for partially flipped Dicke states given in Eq.~\eqref{eq:dicke-twist} the eigenvalues of the constituents $\ket{m}_\rma \ket{{-}m}_\rmb$ add up to zero.
$\ket{m}_\rma \ket{{-}m}_\rmb$ are eigenstates of $J_{z,\rma}$ and $J_{z,\rmb}$, and when acting on them with $J_{x, \rma}$ or $J_{x, \rmb},$ the result is a linear combination of $\ket{m{\pm} 1}_\rma \ket{{-}m}_\rmb$ or $\ket{m}_\rma \ket{{-}m{\mp} 1}_\rmb.$
Hence, the state is orthogonal to the initial state.
Therefore, the result of Equation~\eqref{eq:expect-of-m2} must be zero.
Moreover, all expectation values with an odd number of $J_{l,r}$ operators are zero for $l=x,y.$
Finally, the variance is equal to the second moment of the $\mathcal{M}$ operator,
\begin{multline}
    \variance{\mathcal{M}}_\varrho= \expect{\mathcal{M}^2}_\varrho =
    \expect{J_{z,\rma}^2J_{x,\rmb}^2}_\varrho - \null \\
    - \expect{J_{z,\rma}J_{x,\rma}J_{x,\rmb}J_{z,\rmb}}_\varrho - \expect{J_{x,\rma}J_{z,\rma}J_{z,\rmb}J_{x,\rmb}}_\varrho
    + \expect{J_{x,\rma}^2J_{z,\rmb}^2}_\varrho=\\
    =
    \frac{N_\rma N_\rmb N(3N-10+(N_\rma{-}N_\rmb)^2)}{32(N-3)(N-1)}.
\end{multline}
See Appendix~\ref{app:expectation-values} for more details on the computation of the expectation values.

Thus, the error-propagation formula ("epf") for $\mathcal{M}$ and over the flipped Dicke states is
\begin{equation}
    \label{eq:epf-m2}
    \ivariance{b_1}_{\rm epf} = \frac{2N_\rma N_\rmb(N-3)(N+2)^2}{N(N-1)(3N-10+(N_\rma{-}N_\rmb)^2)}.
\end{equation}
The relative precision compared with the achievable bound \eqref{eq:bound-grad-twited-dicke-general} is
\begin{equation}
    \label{eq:relative-error}
    \frac{\ivariance{b_1}_{\rm epf}}{\ivariance{b_1}_{\rm max}}=
    \frac{(N+2)(N-3)(N-2+(N_\rma{-}N_\rmb)^2)}{(N-1)(N-2)(3N-10+(N_\rma{-}N_\rmb)^2)},
\end{equation}
which for equal splitting becomes
\begin{equation}
    \frac{\ivariance{b_1}_{\rm epf}}{\ivariance{b_1}_{\rm max}}=
    \frac{(N+2)(N-3)}{(N-1)(3N-10)}.
\end{equation}
In the $N\rightarrow\infty$ limit we obtain
\begin{equation}
    \frac{\ivariance{b_1}_{\rm epf}}{\ivariance{b_1}_{\rm max}}\approx \frac{1}{3}.
\end{equation}

Splitting the system into two different wells may introduce an imbalance that depends on the partitioning procedure.
To model the partition noise we consider that $N_\rma = N/2 + \delta$ and $N_\rma = N/2 - \delta$.
The variance for $\delta$, the particle imbalance, is commonly modeled by $(\Delta \delta)^2 =N/4$ \cite{Vitagliano2023NumberPhase}, hence, we consider $(N_\rma-N_\rmb) \propto \sqrt{N}$.
In Eq.~\eqref{eq:relative-error}, the term $(N_\rma-N_\rmb)^2$ can be substituted by $\alpha N$, where the parameter $\alpha$ quantifies the partition noise. 
We obtain
\begin{equation}
    \frac{\ivariance{b_1}_{\rm epf}}{\ivariance{b_1}_{\rm max}}=\frac{(N+2)(N-3)(N-2+\alpha N)}{(N-1)(N-2)(3N-10+\alpha N)},
\end{equation}
which in the $N\rightarrow\infty$ limit becomes
\begin{equation}
    \frac{\ivariance{b_1}_{\rm epf}}{\ivariance{b_1}_{\rm max}}\approx \frac{1+\alpha
    }{3+\alpha}.\label{eq:alpha_ratio}
\end{equation}
For an increasing $\alpha$, the right-hand side of Eq.~\eqref{eq:alpha_ratio} increases. E.g., for $\alpha=1,$ the right-hand side of Eq.~\eqref{eq:alpha_ratio} is $1/2$. As $\alpha$ increases the relative difference is reduced between the error obtained in Eq.~\eqref{eq:epf-m2} and the Cramér-Rao bound \eqref{eq:bound-grad-twited-dicke-general}. Thus, gradient estimation by measuring Eq.~\eqref{eq:M2} is robust against particle number imbalance, which indicates that our method can be used in experiments.

\section{Conclusions}

We studied gradient metrology with quantum states in a two-sell system. It is known that if we have a Bose-Einstein condensate of two-states atoms, then the precision of gradient metrology has a shot-noise scaling with the particle number. We show that if one applies a simple single-particle operations in one of the wells, then gradient metrology is possible with a maximal precision corresponding to a Heisenberg scaling. We consider measuring the gradient and the homogeneous magnetic field along $x$, $y,$ or $z$ directions, and study the inequalities describing the trade-off between these metrological tasks.

\begin{acknowledgments}
We thank M.~Eckstein, F.~Fr\"owis, I.~L.~Egusquiza, C.~Klempt, J.~Ko\l ody\'nski, M.~W.~Mitchell, M.~Mosonyi, G.~Muga, J.~Siewert, Sz.~Szalay, K.~\.Zyczkowski, G.~Vitagliano, and D.~Virosztek  for discussions.
We acknowledge the support of the  EU (COST Action CA15220), the Spanish MCIU (Grant No.~PCI2018-092896), the Spanish Ministry of Science, Innovation and Universities and the European Regional Development Fund FEDER through Grant No.~PGC2018-101355-B-I00 (MCIU/AEI/FEDER, EU) and through Grant No.~PID2021-126273NB-I00 funded by MCIN/AEI/10.13039/501100011033 and by "ERDF A way of making Europe", the Basque Government (Grant No.~IT1470-22), and the National Research, Development and Innovation Office NKFIH (Grant No.~K124351, No.~K124152, No.~KH129601, 2019-2.1.7-ERA-NET-2021-00036, Advanced Grant No. 152794).
We thank the "Frontline" Research Excellence Programme of the NKFIH (Grant No.~KKP133827).~This project has received funding from the European Union's Horizon 2020 Research and Innovation Programme under Grant Agreement No.~731473 and 101017733 (QuantERA MENTA, Quant\-ERA QuSiED).
We thank Project No.~TKP2021-NVA-04, which has been implemented with the support provided by the Ministry of Innovation and Technology of Hungary from the National Research, Development and Innovation Fund, financed under the TKP2021-NVA funding scheme.
We thank the Quantum Information National Laboratory of Hungary.
G.~T.~is thankful for a  Bessel Research Award from the Humboldt Foundation.
M.~G.~thanks the support of the project PID2023-152724NA-I00, with funding from MCIU/AEI/10.13039/501100011033 and FSE+.
M.~G.~aknowledges the support of MCIN/AEI/10.13039/501100011033 fund and the European Union ‘NextGenerationEU’ PRTR fund [RYC2021-031094-I], of the Ministry of Economic Affairs and Digital Transformation of the Spanish Government through the QUANTUM ENIA Project call—QUANTUM SPAIN Project, of the European Union through the Recovery, Transformation and Resilience Plan—NextGenerationEU within the framework of the Digital Spain 2026 Agenda, and of the CSIC Interdisciplinary Thematic Platform (PTI+) on Quantum Technologies (PTI-QTEP+).
M.~G.~is thankful for the support of the project CEX2023-001292-S funded by MCIU/AEI.
\end{acknowledgments}

\appendix
\section{Proof of Observation~\ref{obs:lmax-var-diff}}
\label{app:proof-fig-merit-gra}
\begin{proof}
Our system is split into two wells. Let us first assume that the states  are symmetric in each of the wells.
In this case, we can map the quantum state of the particles in the two wells into the quantum state of two spins with $j_\rma = N_\rma / 2$ and $j_\rmb = N_\rmb / 2$.
Hence, from angular momentum theory, the total angular momentum spin number can have the following values
\begin{equation}
    \label{eq:angular-momentum-expansion}
    j = |j_\rma {-} j_\rmb|,\dots, j_\rma{+}j_\rmb.
\end{equation}
The minimum and maximum eigenvalues of the total angular momentum operator are
\begin{equation}
    \begin{split}
        \label{eq:lmax-lmin-j2}
        \lambda_{\min}(\bm{J}^2) & = \frac{|N_\rma{-}N_\rmb|(|N_\rma{-}N_\rmb|+2)}{4},\\
        \lambda_{\max}(\bm{J}^2) & = \frac{N(N+2)}{4},
    \end{split}
\end{equation}
where $\lambda_{\min}(X)$ and  $\lambda_{\max}(X)$ denote the smallest and largest eigenvalue of $X,$ respectively.
If we expand the square of the total angular momentum with the angular momentum components of the two ensembles `a' and `b', we obtain
\begin{equation}
    \label{eq:sum-of-var}
    \begin{split}
        \bm{J}^2 & = \sum_{l= x, y, z} (J_{l,\rma}{+}J_{l,\rmb})^2 =
        \sum_{l=x, y, z} (J_{l,\rma}^2 + J_{l,\rmb}^2 +
        2 J_{l,\rma} J_{l,\rmb}) \\
        & = \frac{N_\rma(N_\rma +2)}{4} + \frac{N_\rmb(N_\rmb +2)}{4} + 2 \sum_{l= x, y, z} J_{l,\rma} J_{l,\rmb}.
    \end{split}
\end{equation}
Thus, finding the largest and smallest eigenvalues of $\bm{J}^2$ means maximizing and minimizing, respectively, the eigenvalues of the sum of the correlations $\expect{J_{l,\rma}J_{l,\rmb}}$. 
States maximizing $\expect{\bm{J}^2}$ are living in the $j=(N_\rma+N_\rmb)/2$ subspace, while states minimizing $\expect{\bm{J}^2}$ are living in the $j=|N_\rma-N_\rmb|/2$ subspace.

Hence, we can find that the largest eigenvalue for
\begin{multline}
    \label{eq:sum-of-var-diff}
    \sum_{l= x, y, z} (J_{l,\rma}{-}J_{l,\rmb})^2\\
    = \frac{N_\rma(N_\rma +2)}{4} + \frac{N_\rmb(N_\rmb +2)}{4} - 2 \sum_{l= x, y, z} J_{l,\rma} J_{l,\rmb}
\end{multline}
finding the smallest eigenvalue for the sum of the correlations ${J_{l,\rma}J_{l,\rmb}}$.
We conclude that the states with the smallest eigenvalue for $\bm{J}^2$ maximize the eigenvalue of the operator defined in Eq.~\eqref{eq:sum-of-var-diff}.
Thus, from Eq.~\eqref{eq:sum-of-var} and for such states,
\begin{multline}
    \Big\langle2 \sum_{l= x, y, z} J_{l,\rma} J_{l,\rmb}\Big\rangle\\
    = \frac{|N_\rma{-}N_\rmb|(|N_\rma{-}N_\rmb|+2)}{4} - \frac{N_\rma(N_\rma +2)}{4} - \frac{N_\rmb(N_\rmb +2)}{4}.
\end{multline}
Hence,
\begin{multline}
    \lambda_{\max}\Big(\sum_{l= x, y, z} (J_{l,\rma}{-}J_{l,\rmb})^2\Big) \\
    = \frac{N_\rma(N_\rma +2)}{2} + \frac{N_\rmb(N_\rmb +2)}{2} - \frac{|N_\rma{-}N_\rmb|(|N_\rma{-}N_\rmb|+2)}{4}.
\end{multline}
Using that $|N_\rma{-}N_\rmb|=N_\rma+ N_\rmb -2\min(N_\rma, N_\rmb)$ and $N=N_\rma + N_\rmb$, we arrive at
\begin{equation}
    \lambda_{\max}\Big(\sum_{l= x, y, z} (J_{l,\rma}{-}J_{l,\rmb})^2\Big) = \frac{N(N+2)}{4} + \min(N_\rma, N_\rmb).
    \label{eq:jbound}
\end{equation}

Thus, maximizing and minimizing $\expect{\sum_{l= x, y, z} (J_{l,\rma}{-}J_{l,\rmb})^2}$ means minimizing and maximizing, respectively,  the sum of the correlations $\expect{J_{l,\rma}J_{l,\rmb}}$. 
Clearly, states maximizing $\expect{\sum_{l= x, y, z} (J_{l,\rma}{-}J_{l,\rmb})^2}$ are living in the $j=|N_\rma-N_\rmb|/2$ subspace, while states minimizing it are living in the $j=(N_\rma+N_\rmb)/2$ subspace.

Let us now consider the case that the state of the particles in the two wells are not in the symmetric subspace. Let us consider the case when the state is in the eigenspace of
\begin{equation}
J_{x,l}^2+J_{y,l}^2+J_{z,l}^2\label{eq:Jxyz}
    \end{equation}
    with an eigenvalue $j_l(j_l+1)$ for $l=a,b$ and $j_l$ can be smaller than $N_l.$ Then, analogously to Eq.~(\ref{eq:jbound}), we obtain that
\begin{multline}
    \lambda_{\max}\Big(\sum_{l= x, y, z} (J_{l,\rma}{-}J_{l,\rmb})^2\Big) \\
    =(j_\rma+j_\rmb)(j_\rma+j_\rmb+2) + 2\min(j_\rma,j_\rmb),
\end{multline}
which is smaller than or equal to the right-hand side of Eq.~\eqref{eq:jbound} giving the upper bound for $j_\rma=N_\rma/2$ and $j_\rmb=N_\rmb/2.$ Thus, the largest eigenvalue is maximal if $j_\rma=N_\rma/2$ and $j_\rmb=N_\rmb/2.$

Let us now consider a general quantum state $\varrho$ of the two wells. 
Let us define
\begin{equation}
\varrho_{j_\rma,j_\rmb} = P_{j_\rma,j_\rmb} \varrho P_{j_\rma,j_\rmb},
\end{equation}
where $P_{j_\rma,j_\rmb}$ projects to the eigenspace of Eq.~\eqref{eq:Jxyz}  with an eigenvalue $j_l(j_l+1)$ for $l=\rma,\rmb.$
Then, it is easy to see that with our angular momentum measurements our state is not distinguishable from the state
\begin{equation}
\varrho=\sum_{j_\rma,j_\rmb} p_{j_\rma,j_\rmb} \varrho_{j_\rma,j_\rmb},\label{eq:sumrhojajab}
\end{equation}
thus it is sufficient to consider quantum states of the form given in Eq.~\eqref{eq:sumrhojajab}.
Clearly, we obtain the bound
    \begin{equation}
    \Big\langle \sum_{l= x, y, z} (J_{l,\rma}{-}J_{l,\rmb})^2 \Big\rangle \le \frac{N(N+2)}{4} + \min(N_\rma, N_\rmb),\label{eq:jbound2}
\end{equation}
where the bound can be saturated by some state. Thus, in order to obtain the bound, it was sufficient to consider the case when the state of each well is symmetric.

Finally, we have that
\begin{equation}
\sum_{l= x,y,z} \qfi[\varrho, J_{l,\rma}{-}J_{l,\rmb}] \leqs 4 \lambda_{\max} \Big(\sum_{l= x, y, z} (J_{l,\rma}{-}J_{l,\rmb})^2\Big) .
\end{equation}
Hence, Eq.~\eqref{eq:fig-merit-gra} follows.
\end{proof}

\section{Useful expectation values for Dicke states and flipped Dicke states}
\label{app:expectation-values}

Since we have
\begin{equation}
[R_{z,\rmb}(\theta),J_z]=0
\end{equation}
for all $\theta$, the partially flipped Dicke state is an eigenstate of $J_z$ with an eigenvalue 0, similarly to the Dicke state.

Moreover,
\begin{equation}
[R_{z,\rmb}(\theta),J_{l,a}]=0
\end{equation}
holds, while we have
\begin{equation}
\{R_{z,\rmb}(\theta),J_{l,b}\}=0
\end{equation}
for $l=x,y.$
Hence, we can obtain the relevant expectation values for the flipped Dicke state from the expectation values for the original Dicke state.
First, all first moments of the angular momentum operators remain zero.
Moreover, since the Dicke state is an eigenstate of $J_z^2+J_x^2+J_y^2$ with eigenvalue $N(N+2)/4$, the flipped Dicke state is also an eigenstate of $J_z^2 + (J_{x,\rma}{-}J_{x,\rmb})^2 + (J_{y,\rma}{-}J_{y,\rmb})^2$, and therefore,
\begin{equation}
    \expect{J_z^2} + \expect{(J_{x,\rma}{-}J_{x,\rmb})^2} + \expect{(J_{y,\rma}{-}J_{y,\rmb})^2}=\frac{N(N+2)}{4}.
\end{equation}
By expanding the terms above we arrive at
\begin{subequations}
\begin{align}
        \expect{J_{z,\rma}^2}+2\expect{J_{z,\rma}J_{z,\rmb}}+\expect{J_{z,\rmb}^2} &= 0, \label{eq:Jz2_decomposed}\\
        \expect{J_{x,\rma}^2}-2\expect{J_{x,\rma}J_{x,\rmb}}+\expect{J_{x,\rmb}^2}& = \frac{N(N+2)}{8}, \\
        \expect{J_{y,\rma}^2}-2\expect{J_{y,\rma}J_{y,\rmb}}+\expect{J_{y,\rmb}^2}& = \frac{N(N+2)}{8}.
\end{align}
\end{subequations}

At this point, we can compute the expectation values  $\expect{J_{z,\rma}^2}$ and $\expect{J_{z,\rmb}^2}$ over one of the wells based on the definition of the state  $\ket{\widetilde{\rm D}_N}$ given in Eq.~\eqref{eq:dicke-twist} as
\begin{equation}
    \expect{J_{z,r}^2}= \sum_{m=-N_{\rma}/2}^{N_\rma/2}c_m^2 m^2= \frac{N_\rma N_\rmb}{4(N-1)}\label{eq:Jzr2}
\end{equation}
for $r=\rma,\rmb,$
where $c_m$ is given in Eq.~\eqref{eq:cm}.
Note that $\expect{J_{z,\rma}^2}=\expect{J_{z,\rmb}^2}$ even if $N_\rma=N_\rmb.$
We obtain the same values for the unpolarized Dicke state $\ket{{\rm D}_N},$ since certain terms cancel each other. 
Then, from Eq.~\eqref{eq:Jz2_decomposed} and \eqref{eq:Jzr2} follows
\begin{equation}
    \label{eq:jzajzb-dicke}
    \expect{J_{z,\rma}J_{z,\rmb}}=-\frac{N_\rma N_\rmb}{4(N-1)}.
\end{equation}

Each subsystem belongs to the symmetric subspace,
\begin{equation}
    \expect{\bm{J}^2_r}=\frac{N_r(N_r + 2)}{4}
\end{equation}
for $r=\rma,\rmb.$
Hence, we obtain the following relation
\begin{equation}
    \expect{J_{l,r}^2} = \frac{N_r(N_r + 2)}{8} - \frac{\expect{J_{z,r}^2}}{2}
    =  \frac{N_r(N(N_r +1)-2)}{8(N-1)},
\end{equation}
for $l=x,y$.
We also obtain
\begin{equation}
    \label{eq:jlajlb-dicke}
    \expect{J_{l,\rma}J_{l,\rmb}}=-\frac{N_\rma N_\rmb N}{8(N-1)}
\end{equation}
for $l=x,y$.

Finally, we can compute the expectation value of the total angular momentum operator as
\begin{equation}
    \label{eq:total-angular-momentum-flipped-dicke}
    \expect{\bm{J}^2}=\frac{N(N+(N_\rma{-}N_\rmb)^2-2)}{4(N-1)}\approx \frac{N}{4},
\end{equation}
where the last approximation is for $N\gg 1$ and $N_\rma - N_\rmb$ constant.

For Sec.~\ref{sec:Measurement_operator_based_on_moments}, we need fourth moments. First we obtain
\begin{equation}
    \expect{J_{l,r}^2J_{z,r'}^2}=\frac{N_r(N_r + 2)}{8}-\frac{\expect{J_{z,r}^2J_{z,r'}^2}}{2}\label{eq:lrlrprime}
\end{equation}
for $l=x,y$ and $r,r'=\rma,\rmb$ and $r\ne r'.$
Then, we need
\begin{equation}
   \expect{J_{z,r}^2J_{z,r'}^2}=\sum_{m=-N_{\rma}/2}^{N_\rma/2}c_m^2m^4=\frac{N_\rma N_\rmb(3N_\rma N_\rmb - 2 N)}{16(N-3)(N-1)}.\label{eq:Jza2Jzb2}
\end{equation}
By plugging in Eq.~\eqref{eq:lrlrprime} into Eq.~\eqref{eq:Jza2Jzb2}, we arrive at
\begin{equation}
    \expect{J_{l,r}^2J_{z,r'}^2}=\frac{N_r(N_r + 2)}{8}-\frac{N_\rma N_\rmb(3N_\rma N_\rmb - 2 N)}{32(N-3)(N-1)}.
\end{equation}

On the other hand, since both the Dicke state and the flipped Dicke state are eigenstates of $J_z$ with eigenvalue zero, we have that
\begin{subequations}
\begin{align}
    \expect{J_{z,\rma}J_{x,\rma}J_{x,\rmb}(J_{z,\rma}+J_{z,\rmb})} &= 0,\label{eq:fourthorderzxzza}\\
    \expect{(J_{z,\rma}+J_{z,\rmb})J_{x,\rma}J_{x,\rmb}J_{z,\rmb}} &= 0.\label{eq:fourthorderzxzzb}
\end{align}
\end{subequations}
Combining Eqs.~\eqref{eq:fourthorderzxzza} and \eqref{eq:fourthorderzxzzb}, we arrive at
\begin{equation}
    \expect{J_{z,\rma}J_{x,\rma}J_{x,\rmb}J_{z,\rmb}} = - \expect{J_{z,\rma}J_{x,\rma}J_{z,\rma}J_{x,\rmb}}
    = \expect{J_{x,\rma}J_{z,\rma}J_{z,\rmb}J_{x,\rmb}}.\label{eq:JzxxzJzxzx}
\end{equation}

Finally, we use the following facts.
The angular momentum component $J_{x,r}$ can be decomposed as $2J_{x,r}=J_{+,r}+J_{-,r}$, where
\begin{equation}
\begin{split}
J_{+,r}&=J_{x,r}+iJ_{y,r},\\
J_{-,r}&=J_{x,r}-iJ_{y,r}.
\end{split}
\end{equation}
Only the terms $J_{+,\rma}J_{-,\rmb}$ and $J_{-,\rma}J_{+,\rmb}$ contribute to the expectation value in Eq.~\eqref{eq:JzxxzJzxzx}.
Note that $J_{\pm,\rma}J_{\pm,\rmb}$ sends the states from $\ket{m}_\rma\ket{{-}m}_\rmb$ to $\ket{m{\pm}1}_\rma\ket{{-}m{\pm}1}_\rmb,$ that is, to a state orthogonal to the initial state.
Because of the symmetry of the coefficients of the Dicke state and the flipped Dicke state the terms with $J_{+,\rma}J_{-,\rmb}$ and $J_{-,\rma}J_{+,\rmb}$ are equal to each other.
Hence, we obtain
\begin{align}
    &2\expect{J_{z,\rma}J_{x,\rma}J_{x,\rmb}J_{z,\rmb}} = \expect{J_{z,\rma}J_{+,\rma}J_{-,\rmb}J_{z,\rmb}}\nonumber\\
    &\quad\quad= \sum_{m=-N_{\rma}/2}^{N_\rma/2}c_m^2  m(m-1)(N_\rma/2 + m)(N_\rmb/2 + m)\nonumber\\
    &\quad\quad=\frac{N_\rma N_\rmb(N_\rma - 2)(N_\rmb - 2)N}{16(N-3)(N-1)}.
\end{align}

\section{Calculation of \texorpdfstring{$\qfi[\varrho, J_{l,\rma}{+}J_{l,\rmb}]$}{QFI-hom.} and \texorpdfstring{$\qfi[\varrho, J_{l,\rma}{-}J_{l,\rmb}]$}{QFI-gra.} for GHZ state, Dicke state and their partially flipped analogs}
\label{app:polytope-expectation-values}

GHZ states \eqref{eq:ghz}, Dicke states \eqref{eq:dicke}, partially flipped GHZ states \eqref{eq:ghz-flipped} and partially flipped Dicke states \eqref{eq:dicke-twist} belong to the product of the symmetric subspaces on each of the wells.
Hence, the total angular momentum over each of the wells must be conserved
\begin{equation}
    \label{eq:total-each-well}
    \expect{J_{x,r}^2} + \expect{J_{y,r}^2} + \expect{J_{z,r}^2} = \frac{N_r(N_r+2)}{4},
\end{equation}
for $r=\rma,\rmb.$ 
They are eigenstates of $J_z$ too.
As consequence, they are invariant under the swap of the $x$ and $y$ axes.

Moreover, they are not only unpolarized states as a whole but on each of the wells too, \ie all first moments of the following angular momentum operators are zero, $\expect{J_{l,r}} = 0.$
Hence, in order to compute the QFIs, we have to compute the second moments only.
We also have that 
\begin{equation}
    \label{eq:sum2-expand-ab}
    \expect{(J_{l,\rma}{\pm}J_{l,\rmb})^2} = \expect{J_{l,\rma}^2}\pm 2\expect{J_{l,\rma}J_{l,\rmb}}+\expect{J_{l,\rmb}^2}.
\end{equation}
In Appendix~\ref{app:expectation-values}, we already computed the second moments for the unpolarized Dicke state and the partially flipped Dicke state, Eqs.~\eqref{eq:jzajzb-dicke} and \eqref{eq:jlajlb-dicke}.

With all these relations and expectations values, we complete the Table~\ref{tab:second-moments}.
\begin{table}
    \begin{tabular}[t]{l|cccc}
        \hline\hline
        & $\ket{{\rm GHZ}}$ & $\ket{\widetilde{\rm GHZ}}$ & $\ket{{\rm D}_N}$ & $\ket{\widetilde{\rm D}_N}$ \\[0.25em]
        \hline
        $\qfi[\varrho,J_{z}]$   & $N^2$ & $(N_\rma - N_\rmb)^2$ & 0 & 0 \\[0.5em]
        $\qfi[\varrho,J_{l}]$   & $N$ & $N$ & $\frac{N(N+2)}{2}$ & $\{\mathcal{F}\}$ \\[0.5em]
        $\qfi[\varrho,J_{z,\rma}{-}J_{z,\rmb}]\;$ & $\;(N_\rma - N_\rmb)^2$ & $N^2$ & $\left\{\frac{4N_\rma N_\rmb}{N-1}\right\}$ & $\left\{\frac{4N_\rma N_\rmb}{N-1}\right\}$ \\[0.5em]
        $\qfi[\varrho,J_{l,\rma}{-}J_{l,\rmb}]$ & $\{N\}$ & $\{N\}$& $\{\mathcal{F}\}$ & $\frac{N(N+2)}{2}$ \\[0.5em]

        \hline\hline
    \end{tabular}
    \caption{
        \label{tab:second-moments}
        QFI for GHZ states, partially flipped GHZ states, Dicke states and partially flipped Dicke states, where $l=x,y.$
        Some QFIs are computed directly from the definition of the states, Eqs.~\eqref{eq:ghz}, \eqref{eq:ghz-flipped}, \eqref{eq:dicke-subhilb} and \eqref{eq:dicke-twist}.
        (curly-braces) The remaining ones are obtained using the relations Eqs.~\eqref{eq:total-each-well} and \eqref{eq:sum2-expand-ab}.
        $\mathcal{F}$ is shown in the main text in Eq~\eqref{eq:qfi-jl-pdicke} due to lack of space.
    }
\end{table}
We have for Dicke and partially flipped Dicke states that
\begin{equation}
    \label{eq:qfi-jl-pdicke}
    \mathcal{F} = \frac{N(N+(N_\rma - N_\rmb)^2 - 2)}{2(N-1)} \approx \frac{N}{2},
\end{equation}
where the last approximation is for $N\gg 1$ and $N_\rma - N_\rmb$ constant.
Equation~\eqref{eq:total-angular-momentum-flipped-dicke} shows a similar result from which the equation \eqref{eq:qfi-jl-pdicke} could also be derived.
Note that the partially flipped Dicke state has similar expectation values if one exchanges the $x$- and $y$-axes.
For the case shown on Figure~\ref{fig:polytopes}, $N_\rma = N_\rmb = 4$.

On the Table~\ref{tab:second-moments}, the first two rows of the partially-flipped Dicke state $\ket{\widetilde{\rm D}_N}$ correspond to four times the variance of $\variance{J_z}$ and $\variance{J_l}$, respectively. 
Summing all the three variances one reaches to $N/4$ for large systems, which clearly violates the entanglement criterion for singlet-like states  \cite{Toth2004Entanglement,Toth2007Optimal,Toth2010Generation,Vitagliano2011Spin}
\begin{equation}
    [\Delta (J_x^{\rm a}+J_x^{\rm b})]^2+[\Delta (J_y^{\rm a}+J_y^{\rm b})]^2+[\Delta (J_{\rm z}^{\rm a}+J_{\rm z}^{\rm b})]^2\ge \frac{N_{\rm a}}2+\frac{N_{\rm b}}2,\label{eq:threevarent}
\end{equation}
Clearly, for Dicke states the following entanglement criterion is violated \cite{Lange2018Entanglement,Vitagliano2023NumberPhase}
\begin{equation}
    [\Delta (J_x^{\rm a}-J_x^{\rm b})]^2+[\Delta (J_y^{\rm a}-J_y^{\rm b})]^2+[\Delta (J_{\rm z}^{\rm a}+J_{\rm z}^{\rm b})]^2\ge \frac{N_{\rm a}}2+\frac{N_{\rm b}}2.
\end{equation}
In Ref.~\cite{Vitagliano2023NumberPhase}, a similar relation is used to detect entanglement between two halves of a Dicke state of two-state atoms in a Bose-Einstein condensate. Unlike, the partially-flipped Dicke state $\ket{\widetilde{\rm D}_N},$ the partially-flipped GHZ state $\ket{\widetilde{\rm GHZ}_N}$ does not violate Eq.~\eqref{eq:threevarent}. However, for $N_{\rm a}=N_{\rm b},$ it saturates the entanglement criterion, thus the three collective variances are still small for that state.

\section{Number of independent Hermitian operators that commute with \texorpdfstring{$J_y$}{Jy} when the product of two spins \texorpdfstring{$j_\rma$}{ja} and \texorpdfstring{$j_\rmb$}{jb} is considered.}
\label{app:independent-hermitian-operators}

{\it Observation S1.}
    The total number of independent Hermitian operators, $M_k$, that commute with $J_y$ for a combined system of two subsystems, $j_\rma=N_\rma / 2$ and $j_\rmb = N_\rmb / 2$, is  
    \begin{equation}
        \label{eq:independent-hermitian}
        K^*_{N_{\min}, N_{\max}} = 1 + N_{\min} + \frac{N_{\min}(1- N_{\min}^2)}{3} + (1 + N_{\min})^2 N_{\max}
    \end{equation}
    where $N_{\min} \equiv \min(N_\rma,N_\rmb)$ and $N_{\max} \equiv \max(N_\rma,N_\rmb)$.
    If we impose that $\variance{M_k}_\varrho=0$ for Dicke states given in Eq.~\eqref{eq:dicke}, the number is smaller
    \begin{equation}
        \label{eq:independent-hermitian-reduced}
        K_{N_{\min}, N_{\max}} = \frac{K^*_{N_{\min}, N_{\max}} + N_{\min} + 1}{2},
    \end{equation}

\begin{proof} We will consider only the symmetric states in the two subspaces a and b in order to reduce the number of operators.
As in Eq.~\eqref{eq:angular-momentum-expansion}, the whole Hilbert space can be considered as adding $j_\rma=N_\rma/2$ and $j_\rmb=N_\rmb/2$ angular momentum subspaces. Thus, the dimension of the space is $(N_\rma+1)(N_\rmb+1).$ Equivalently, from quantum angular momentum theory, the dimension is 
$\sum_{j=j_{\min}}^{j_{\max}} 2j+1$, where there are
$j_{\max}-j_{\min}+1=\min(N_\rma, N_\rmb)+1$ independent total angular momentum subspaces and $j_{\max}=j_\rma+j_\rmb$ and $j_{\min}=|j_\rma-j_\rmb|$.

It is easy to see that the degeneracy of the eigenvalues of $J_y$ follow
\begin{equation}
    d(m_y)=\begin{cases}
        \min(N_\rma, N_\rmb)+1 & \textnormal{if $|m_y| \leqs j_{\min}$,}\\
        \frac{N_\rma+N_\rmb}{2} + 1 - |m_y| & \textnormal{otherwise}.
    \end{cases}
\end{equation}
Hence, in a degenerate subspace there are $d(m_y)^2$ independent Hermitian operators which commute with the original operator, $d(m_y)$ diagonal terms and $d(m_y)[d(m_y)-1]$ real and complex off-diagonal terms.

Thus, the total number of independent Hermitian operators is
\begin{multline}
    \label{eq:sum-number-independent-hermitian}
    K^{*} = \sum_{m_y=-j_{\min}}^{j_{\min}} [\min(N_\rma, N_\rmb)+1]^2 + \\ + 2\sum_{m_y=j_{\min}+1}^{j_{\max}} \left(\tfrac{N_\rma+N_\rmb}{2} + 1 - m_z\right)^2,
\end{multline}
where the factor of two appears in front of the the second sum taking into account  a summation for $m_y = -j_{\max},\dots,-(j_{\min}+1)$. Finally, using that 
\begin{equation}
2|j_\rma-j_\rmb|=\max(N_\rma,N_\rmb)-\min(N_\rma,N_\rmb),
\end{equation}
we arrive at
\begin{multline}
    K^{*} = \frac{1+\min(N_\rma,N_\rmb)}{3}\bigl\{3+\min(N_\rma,N_\rmb)-\min(N_\rma,N_\rmb)^2+\\+3\max(N_\rma,N_\rmb)[1+\min(N_\rma,N_\rmb)]\bigr\},
\end{multline}
which is equivalent to Eq.~\eqref{eq:independent-hermitian}.

Our system consisting of a partially flipped Dicke state given in Eq.\eqref{eq:dicke-twist}, has non-zero overlap with eigenstates of eigenvalues $m_y=j_{\max}, j_{\max}-2, \dots, -j_{\max}+2, -j_{\max},$ which is the case also for the Dicke state.
In that case, only Hermitian operators that act on such subspace may have $\variance{M_k}_{\varrho}\,{\neq}\,0$.
Therefore, the sum in Eq.~\eqref{eq:sum-number-independent-hermitian} reduces to 
\begin{multline}
    K^{*} = \sum_{\mu_y=-j_{\min}/2}^{j_{\min}/2} (\min(N_\rma, N_\rmb)+1)^2 + \\ + 2\sum_{\mu_y=j_{\min}/2+1}^{j_{\max}/2} (\tfrac{N_\rma+N_\rmb}{2} + 1 - 2\mu_z)^2,
\end{multline}
where we introduced a new summation variable $\mu_y = 2m_y$ and changed the limits accordingly.
Equation \eqref{eq:independent-hermitian-reduced} follows.
\end{proof}

These two numbers, Eqs.~\eqref{eq:independent-hermitian} and \eqref{eq:independent-hermitian-reduced}, scale as $\mathcal{O}(N_{\min}^2N_{\max})$ for large $N_{\min}$ and $N_{\max}$. 
Hence, the amount of computations required for not so-large systems increases rapidly.

\bibliography{Bibliography2_DickeMetrology}

\end{document}